\documentclass[prb,reprint,superscriptaddress]{revtex4-1}

\usepackage{graphicx}  %
\usepackage{bm}        %
\usepackage{amssymb}   %

\begin{document}

\title{High-harmonic generation from few layer hexagonal boron nitride:\\
        evolution from the monolayer to the bulk response}

 \author{Guillaume Le Breton}
  \affiliation{D\'epartement de Physique, \'Ecole Normale Sup\'erieure de Lyon, \\Universit\'e Claude Bernard Lyon 1, 46 All\'ee d'Italie, Lyon Cedex 07, France}
  \affiliation{Max Planck Institute for the Structure and Dynamics of Matter, Luruper Chaussee 149, 22761 Hamburg, Germany}

 \author{Angel Rubio}
  \email{angel.rubio@mpsd.mpg.de}
  \affiliation{Max Planck Institute for the Structure and Dynamics of Matter, Luruper Chaussee 149, 22761 Hamburg, Germany}

 \author{Nicolas Tancogne-Dejean}
  \email{nicolas.tancogne-dejean@mpsd.mpg.de}
  \affiliation{Max Planck Institute for the Structure and Dynamics of Matter, Luruper Chaussee 149, 22761 Hamburg, Germany}

\begin{abstract}

Two-dimensional materials offer a versatile platform to study high-harmonic generation (HHG), encompassing as limiting cases bulk-like and atomic-like harmonic generation [Tancogne-Dejean and Rubio, Science Advance \textbf{4}, eaao5207 (2018)].
Understanding the high-harmonic response of few-layer semiconducting systems is important, and might open up possible technological applications.
Using extensive first-principle calculations within a time-dependent density functional theory framework, we show how the in-plane and out-of-plane nonlinear non-perturbative response of two-dimensional materials evolve from the monolayer to the bulk. We illustrate this phenomenon for the case of multilayer hexagonal BN layered systems.
Whereas the in-plane HHG is found not to be strongly altered by the stacking of the layers, we found that the out-of-plane response is strongly affected by the number of layers considered. This is explained by the interplay between the induced electric field, resulting from the by electron-electron interaction, and the interlayer delocalization of the wave-functions contributing most to the HHG signal. The gliding of a bilayer is also found to affect the high-harmonic emission.
Our results will have important ramifications for the experimental study of monolayer and few-layer two-dimensional materials beyond the case of hexagonal BN studied here as the result we found are generic and applicable to all 2D semiconducting multilayer systems.
\end{abstract}

\maketitle

\section{\label{Sec: Introduction} Introduction}

Understanding the non-linear non-perturbative response of two-dimensional (2D) materials might open up new applications in the emerging fields of strong-field optoelectronics\cite{schiffrin_optical-field-induced_2012,sivis_tailored_2017} and petahertz electronics.\cite{garg_multi-petahertz_2016}
Whereas the two-dimensional materials are at the hearth of a huge scientific activity, with a strong focus on valleytronics \cite{schaibley2016valleytronics,langer2018lightwave}, and the engineering of van-der-Walls heterostructures~\cite{Novoselovaac9439}, little is known on the non-linear non-perturbative response of 2D materials to strong laser fields.
Previous experimental and theoretical works focused on monolayer materials, including graphene\cite{yoshikawa_high-harmonic_2017,cox_plasmon-assisted_2017,chizhova_high-harmonic_2017,dimitrovski_high-order_2017,taucer_nonperturbative_2017,1802.02161,PhysRevB.82.201402,PhysRevB.91.045439,PhysRevB.93.155434,PhysRevB.90.085313,higuchi2017light}, transition metal dichalcogenides\cite{liu_high-harmonic_2016,yoshikawa_high-harmonic_2017,langer2018lightwave}, and monolayer hexagonal boron nitride (hBN)\cite{tancogne-dejean_atomic-like_2018}. 
It was reported experimentally that MoS$_2$ generate harmonics more efficiently than in the bulk \cite{liu_high-harmonic_2016}.
Graphene and MoS$_2$ have also been shown to exhibit particular ellipticity dependence\cite{yoshikawa_high-harmonic_2017}.
These works mostly investigated the in-plane high-harmonic generation (HHG) from 2D materials, whereas some of us recently demonstrated that 2D materials can generate atomic-like high-order harmonics if driven by an out-of-plane polarized laser field\cite{tancogne-dejean_atomic-like_2018}.
By atomic-like HHG, we mean that electrons are promoted to the continuum and follow well-defined trajectories in real space, which can be explained  by the three-step model of HHG in atoms. The energy cutoff $E_c$ of the HHG from a 2D material driven by a laser polarized along the out-of-plane direction was also shown to be $E_c = E_w + 3.17 U_p$, where the work function $E_w$ plays the role of the ionization potential.~\cite{tancogne-dejean_atomic-like_2018}
In the limit of an infinite number of layers, one should recover the properties of a bulk material, which generates solid HHG. It is therefore expected that a transition will occur in between the monolayer case and the bulk case, in which a bulk character will emerge. This transition corresponds to delocalization of the electronic states along the out-of-plane direction, related to the emergence of the electronic bands which are iconic to periodic bulk materials. This study aims at investing in details this transition between atomic-like HHG and solid HHG.\\
Few-layer systems can nowadays be prepared with a very high degree of control~\cite{Novoselovaac9439, D_Cory_group_nature_2018, James_Hone_group_2015,dean2010boron}, and offer a novel playground for engineering tailored electronic and optical properties. It is therefore very desirable to understand how the stacking of layers affects their optical properties in the context of strong-field physics.\\
Behind the possibility of controlling HHG from tailored van der Walls heterostructures lies some more fundamental questions: (i) How the HHG evolves while stacking layers from a monolayer to the bulk? (ii) How many layers are necessary to recover the bulk properties? (iii) How the surface impacts the few-layer nonlinear response? (iv) Does the HHG depend on the details of the stacking sequence between layers, and is it affected by the sliding of the layers? Understanding in details these points will be of major importance for future experimental studies, and possible technological applications. Moreover, identifying clear fingerprints related to a particular stacking, or the number of layers, could open up the possibility of using HHG as a spectroscopical tools for characterizing structurally few-layer systems. 
It is the very purpose of this paper to address these questions, in order to gain deeper insight on the HHG from monolayer and few-layer systems, as well as on surface effects.\\
In this paper, using an \textit{ab initio} approach based on time-dependent density functional theory \cite{runge_density-functional_1984,van_leeuwen_mapping_1999}, we study the HHG from monolayer, few-layer and bulk systems. Our calculations take fully into account the full band-structure of the various systems, and include the electron-electron interactions, which has been shown to be important for the HHG from free-standing monolayer hBN~\cite{tancogne-dejean_atomic-like_2018}.\\
This paper is organized as follow: The methodology is discussed in Sec.~\ref{sec:method}. We then present our results for monolayer, few-layer and bulk hBN in Sec.~\ref{sec:results}. Section~\ref{sec:discussion} discusses the implications of our findings and important points that go beyond the in-plane and out-of-plane HHG responses. Conclusions are summarized in Sec.~\ref{sec:conclusions}.

\section{Methodology}
\label{sec:method}

\subsection{Numerical details}

\begin{figure}[h]
      \includegraphics[width=0.6\columnwidth]{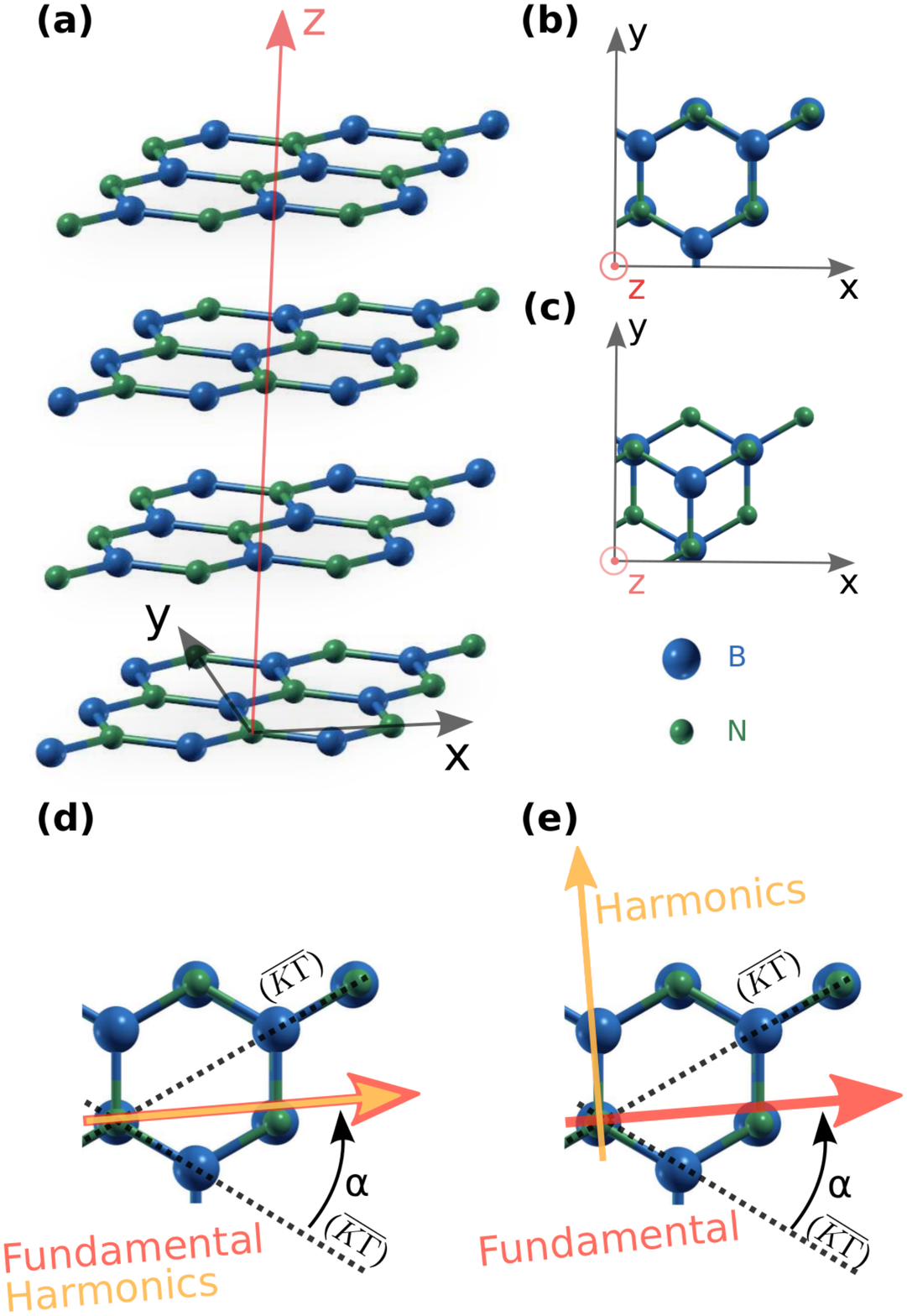} %
    \caption{\label{fig:view_system} (a) Side view and (b) top view of a four-layer hBN slab with AA$'$ stacking as explained in the main text. (c) Another possible stacking, referred below as the AD stacking. Decomposition of the harmonic emission into (d) parallel and  (e) perpendicular contributions, defined with respect to the laser polarization direction. This latter is defined by the angle $\alpha$ with respect to the crystallographic direction $\overline{K \Gamma}$. }
\end{figure}

In order to investigate the effect of the stacking, we selected hexagonal boron nitride as a prototypical material.
Calculations are performed in the framework of time-dependent density functional theory (TDDFT). The time-evolution of the wavefunctions and the evaluation of the total electronic current are done using the real-time real-space Octopus code \cite{andrade_real-space_2015,octopus_PSSb_2006,octopus_sciencedirect_2003,propagator_KS_jcp2004}, within the adiabatic local-density approximation (LDA) \cite{LDA}. Note that the quantitative results being presented here do not depend strongly on the choice  of the exchange-correlation potential and we checked that the PBE functional was leading to very similar results.
The B-N distance is taken as the experimental value\cite{hbn_exp_1952} of 2.73 Bohr and the distance between two consecutive layers is 6.29 Bohr, which corresponds to the equilibrium distance \cite{hbn_configuration}. 
Unless stated differently, we considered throughout this work a AA$'$ stacking, which corresponds to having boron atoms of a layer on top of the nitrogen atoms of the next layer and \textit{vice versa}. This is illustrated in Fig.~\ref{fig:view_system}.
As we will discuss in Sec.~\ref{sec:discussion}, the details of the stacking, and in particular the so-called AD configuration~\cite{hbn_configuration}, lead to a modification of the HHG response of few-layer hBN.
Mixed periodic boundary conditions are used in the in-plane directions and in the out-of-plane direction. We used the primitive (hexagonal) cell containing one boron and one nitrogen atom per layer.

The total size of the simulation box along the out-of-plane direction is 480 Bohr, including absorbing layers of 40 Bohr on each side in order to avoid unphysical reflection of field-accelerated electrons at the border of the simulation box. As absorbing boundaries, we employed the complex absorbing potential (CAP) method\cite{absorbing_boundary_octopus}, with a cap height $\eta=-1$ a.u.. The real-space box is sampled along all directions by a grid spacing of 0.28 Bohr. The 2D Brillouin zone is sampled by a uniform $32\times32$ $\mathbf{k}$-point grid for out-of-plane HHG calculations and by a uniform $42\times42$ $\mathbf{k}$-point grid for in-plane HHG calculations. For the bulk, we employed a grid of $54\times 54\times 27$ $\mathbf{k}$-points.  The size of the box along the out-of-plane direction, the real-space spacing and the number of $\mathbf{k}$-point have been converged with respect to the output harmonic spectra, to give less than 3\% deviation to a more converged reference spectrum. 
The driving laser field is assumed to  be spatially-uniformed, and is described in the velocity gauge to preserve the in-plane periodicity. We considered a laser of 15  fs pulse duration  at full-width half maximum for the out-of-plane case and 30 fs for the in-plane case, with a wavelength centered at 1600 nm (corresponding to a photon energy of 0.77 eV), and a sin-square envelope with a zero carrier-envelope phase (CEP). \\
In order to get proper structural properties, we also performed calculations including the van der Waals (vdW) interaction using the method introduced by Tkatchenko and Scheffler ~\cite{tkatchenko_accurate_2009}. For the electronic dynamics discussed in  Sec.~\ref{sec:results}, we however neglected the vdW interaction, as we found no significant changes including it.\\
The HHG spectrum is computed from the total electronic current $\mathbf{j}(\mathbf{r},t)$ as (atomic units are used throughout this paper)\cite{PhysRevB.77.075330}

\begin{equation}
\label{Eq:hhg_formula}
\mathrm{HHG}(\omega) = \left|\mathcal{F} \left( \frac{\partial}{\partial t} \int_{\Omega} \mathrm{d}^3\mathbf{r}\, \mathbf{j}(\mathbf{r},t)\right) \right|^2,
\end{equation} 
where $\mathcal{F}$ denotes the Fourier transform and $\Omega$ the volume of the simulation box.

Using the aforementioned parameters, we obtained that the bulk has a 4.87 eV gap at the $\mathbf{k}$-point K and finite systems have gaps ranging from 4.50 eV (monolayer) to 4.63 eV (hexalayer). Values for the other systems can be found in Tab.~\ref{tab:comp_BG_WF}, together with the values of their work function.
\begin{table}[h]
\begin{ruledtabular}
\begin{tabular}{l|c|c}
Number& Band gap & Work function \\
 of layers  & (eV) & (eV)  \\
         \hline
1 &  4.50 & 6.11 \\
2 &  5.05 & 6.39 \\
3 &  4.60 & 6.26 \\
4 &  4.54 & 6.13 \\
5 &  4,51  & 6.10 \\
6 &  4.63 & 6.22 \\
bulk & 4.87 & - \\
\end{tabular}
\end{ruledtabular}
\caption{\label{tab:comp_BG_WF}Calculated values of the band gap and the work function, in eV, of the various slabs. The stacking is AA$'$. }
\end{table}

\subsection{Induced electric field}

As one of the key result of this work, we found that the induced electric field plays a central role in the description of the HHG from few-layer hBN, and in particular when the driver is polarized along the out-of-plane direction.
The longitudinal part of the induced electric field is naturally included in our real-time TDDFT simulations as we are propagating the time-dependent Kohn-Sham equations within the adiabatic approximation
\begin{eqnarray}
 i\frac{\partial}{\partial t}\psi_{i}(\mathbf{r},t) = \Big[-\frac{\nabla^2}{2} + v_{\mathrm{ext}}(\mathbf{r},t) + v_{\mathrm{H}}[n(\mathbf{r},t)](\mathbf{r}) \nonumber\\
 + v_{\mathrm{xc}}[n(\mathbf{r},t)](\mathbf{r})\Big]\psi_{i}(\mathbf{r},t).
\end{eqnarray}
In this equation, $i$ refers to both a band and a $\mathbf{k}$-point index, $v_{\mathrm{ext}}$ is the external potential containing both the driving laser field and the ionic potential, $v_{\mathrm{H}}$ is the Hartree potential, and $v_{\mathrm{xc}}$ is the exchange-correlation potential. We omitted here the nonlocal contribution to the external potential from the pseudopotentials for simplicity. In the equation, the laser is described in the velocity gauge, i.e., the corresponding time-dependent potential perturbing the system is $v(t)=\frac{1}{c}\mathbf{A}(t)\ldotp\mathbf{p} + \frac{1}{2c^2}\mathbf{A}^2(t)$.

The longitudinal induced electric field taken into account in our calculations is related to the gradient of the time-variation of the Hartree potential. Indeed, starting from from Gauss' law, and using the linearity of Maxwell equations, we have
\begin{equation}
\mathbf{\nabla}.\mathbf{E}_{\mathrm{ind}}(\mathbf{r},t)=4 \pi n_{\mathrm{ind}}(\mathbf{r},t), 
\label{eq:gauss}
\end{equation}
where $n_{ind}$ denotes the induced electronic density, i.e., the difference between the time-evolved density (at time t) and the groundstate one (at initial time $t_0$), $n_{\mathrm{ind}}(\mathbf{r},t)= n(\mathbf{r},t)-n(\mathbf{r},t_0)$. One easily obtain that 
\begin{eqnarray}
 \mathbf{E}^L_{\mathrm{ind}}(\mathbf{r},t) = \nabla_{\mathbf{r}} \left[ v_{\mathrm{H}}(\mathbf{r},t) - v_{\mathrm{H}}(\mathbf{r},t_0)  \right].
 \label{eq:eind}
\end{eqnarray}

Using this expression, it is therefore possible to compute the induced electric field that it is accounted for in our simulations. It is important to note here even if the external field is spatially uniform, our simulations take into accounts the spatial fluctuations of the induced electric field, which are responsible for the screening of the electric field by surface charges induced by the external field \cite{tancogne-dejean_atomic-like_2018}, as discussed later.

\section{Results}
\label{sec:results}

\subsection{\label{Sec:inplane} In-plane}

\begin{figure*}
 \includegraphics[width=1\columnwidth, angle = -90]{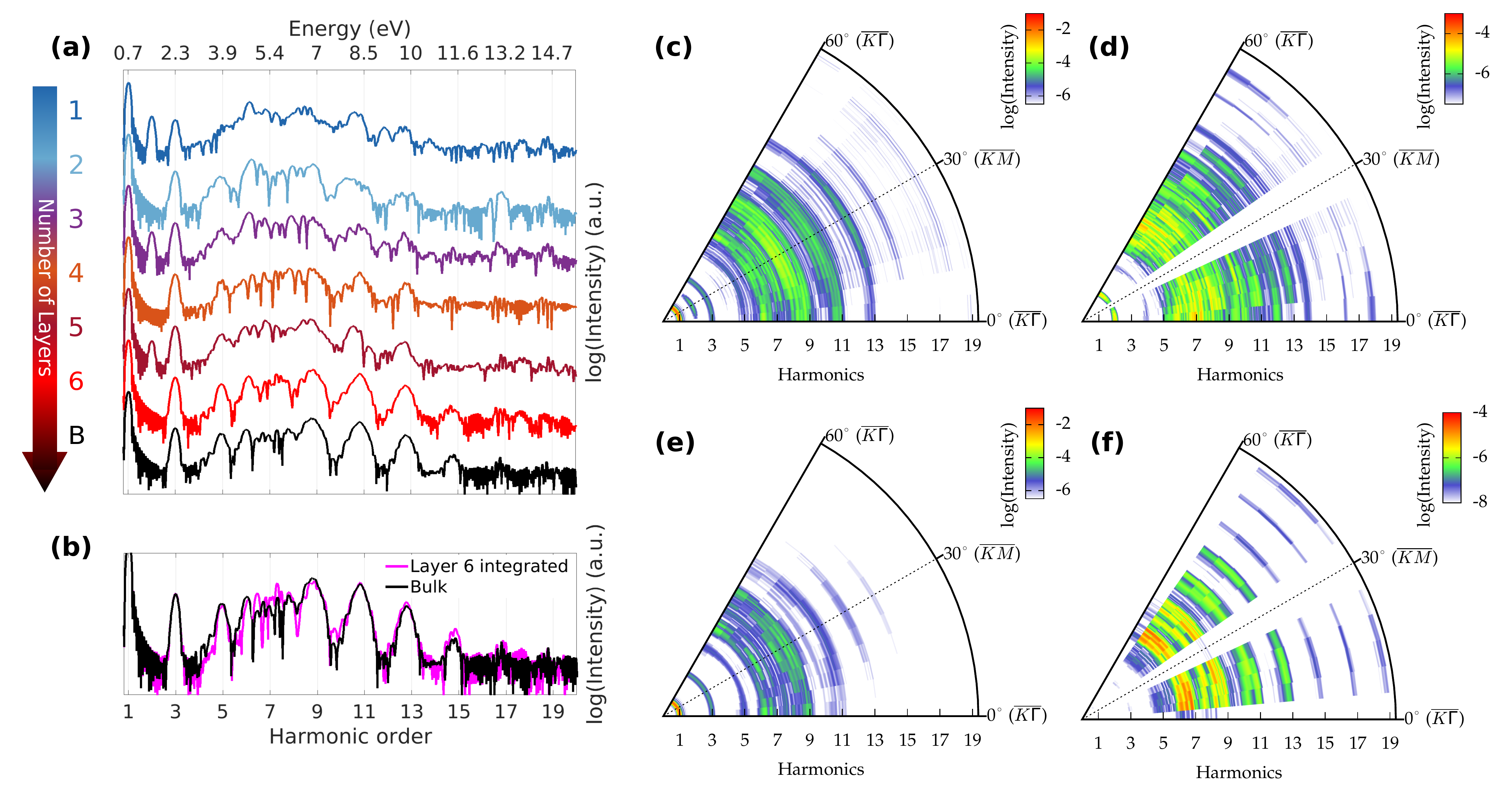} %
    \caption{\label{fig: spectra inplane} (a) Normalized HHG spectra for one to six-layer slabs and for the bulk. The laser polarization is taken along the $\overline{K \Gamma}$ crystallographic direction. (b) Comparison between the bulk HHG spectrum (black curve) and the spectrum obtained from the inner-most layers of a six-layer slab, as explained in the text. (c) Parallel and (d) perpendicular contributions to the HHG spectrum of the monolayer \textit{versus} the in-plane polarization angle, as defined in Fig.~\ref{fig:view_system}. (e-f) Same as (c-d) but for the bilayer case.}
\end{figure*}

We start by analyzing the effect of the layer stacking on the in-plane HHG spectra from few-layer hBN. 
For this, we computed the HHG spectra for one layer up to six hBN layers, as well as for bulk hBN.
Our results, reported in Fig.~\ref{fig: spectra inplane}(a), are obtained for an intensity \textit{in matter} of $7.02 \times 10^{13}$ W.cm$^{-2}$, using the
experimental in-plane refractive index n = 2.65.\cite{h-BN_refractive_index}
In order to compare the HHG spectra for the different number of layers and for the bulk, we normalized the electronic current to the number of layers.
Harmonics below the gap are well determined, as well as higher-order ones, whereas the region close the gap is not perfectly resolved, similar to what was found in previous studies of bulk materials, see for instance Ref.~\onlinecite{tancogne-dejean_impact_2017} and references therein.
As found experimentally in Ref.~\onlinecite{liu_high-harmonic_2016}, the harmonic yield of the monolayer is higher than the yield of few-layer hBN or than the bulk. Moreover we observe that while for bulk hBN the calculated energy cutoff corresponds to the harmonic order 15, harmonics up to the order 19 are obtained in the case of the monolayer. 
Therefore the few-layer systems emit more intense and more harmonics than the bulk counterpart.\\
The parity of the number of layers has a direct impact on the spectra, as even harmonics are visible for a odd number of layers, as expected from simple symmetry considerations.
Indeed, for an even number of layers (with AA$'$ stacking) a slab has inversion symmetry, which prohibits even harmonics.
The contribution to the even harmonics from one layer compensates the one of the next layer, as each pair of layers behaves as a centrosymmetric material. As a result, no even harmonics are obtained for an even number of layers, and a net contribution from only one layer is obtain for an odd number of stacked layers.
This explain why even harmonics are very similar for all the slabs made of an odd number of layers, because these even harmonics arise from the effective contribution of one layer.

While increasing the number of layers, the HHG spectra quickly converge to the one of the bulk material.
However, even for the six-layer slab, some discrepancies still persist with the bulk HHG spectrum. This is easily understood, as the slab contains contributions originating from the surfaces and its finite size, which are absent in the case of the bulk. Increasing the number of layers N, it is clear the relative weight of the surface contribution will slowly decrease as 1/N. To confirm this, we performed a layer-by-layer analysis in  Fig.~\ref{fig: spectra inplane}(b), assuming that each bilayer contributes independently of the others. The total electronic current is first integrated along the out-of-plane direction over the volume of a bilayer, and then used to compute the corresponding HHG spectrum. The HHG spectrum from the inner-most layers of a six-layer slab, as well as the bulk HHG spectrum are plotted in Fig.~\ref{fig: spectra inplane}(b). While the spectral contribution of the external layers are quite different (not shown) from the bulk spectrum, it is clear that HHG spectrum from the inner-most layers is almost identical to the bulk spectrum, showing that few-layer systems very quickly recover the pure bulk character.\\
Due to the hexagonal symmetry of hBN layers, harmonic emission is not only polarized along the polarization direction of the driving field. Following Ref.~\onlinecite{liu_high-harmonic_2016}, we split the harmonic emission into a parallel contribution, which corresponds to the emission along the polarization direction of the driving field, and a perpendicular contribution. This is sketched in Figs.~\ref{fig:view_system}(d-e).\\
In Figs.~\ref{fig: spectra inplane}(c)-(d), we show the anisotropy of the HHG emission for the monolayer h-BN while rotating the laser polarisation in the plane of the material. The mirror plane along the $\overline{KM}$ cristallographic direction is clearly visible in these maps. Similar to the result obtained experimentally in a MoS$_2$ monolayer~\cite{liu_high-harmonic_2016}, the parallel contribution produces odd and even harmonics. At variance with Ref.~\onlinecite{liu_high-harmonic_2016}, the perpendicular contribution contains not only even harmonics, but seems to also contain odd harmonics.  In the bilayer case (Figs.~\ref{fig: spectra inplane}(e-f)), no even harmonics are generated, due to the centro-symmetry of such a bilayer. As a result, odd harmonics are observed in both parallel and perpendicular contributions.

To summarize our findings for an in-plane driving field, we found that i) the monolayer is more efficient to generate HHG than stacked layers and the bulk system, similarly to what was found experimentally in MoS$_2$\cite{liu_high-harmonic_2016}.
ii) The HHG spectrum converges very quickly with the number of layers, and matches well the bulk spectrum for N$\ge6$ layers. 
iii) Even harmonics are generated from slabs with an odd number of layers. These even harmonics depend strongly on the symmetries of the system, see Sec.~\ref{sec:discussion} for a discussion on the effect of the stacking. The intensity of the even harmonics is almost not affected by the number of layers stacked, as they originate from the remaining part of destructive interferences. 

\subsection{\label{Sec:outofplane} Out-of-plane}

We now consider the case of an out-of-plane driving field. As shown in Ref.~\onlinecite{tancogne-dejean_atomic-like_2018}, a free-standing two-dimensional material generates atomic-like harmonics, in which electrons follow semi-classical trajectories.
However, it is clear that this picture cannot hold in the bulk anymore, in which harmonic generation has a different microscopic origin~\cite{huttner_similarities_2016}. We therefore investigate how the out-of-plane HHG response evolves from the monolayer to the bulk, by computing the HHG from few-layer hBN slabs driven by a driving field polarized along the out-of-plan direction.

\begin{figure}

 \includegraphics[width=1\columnwidth]{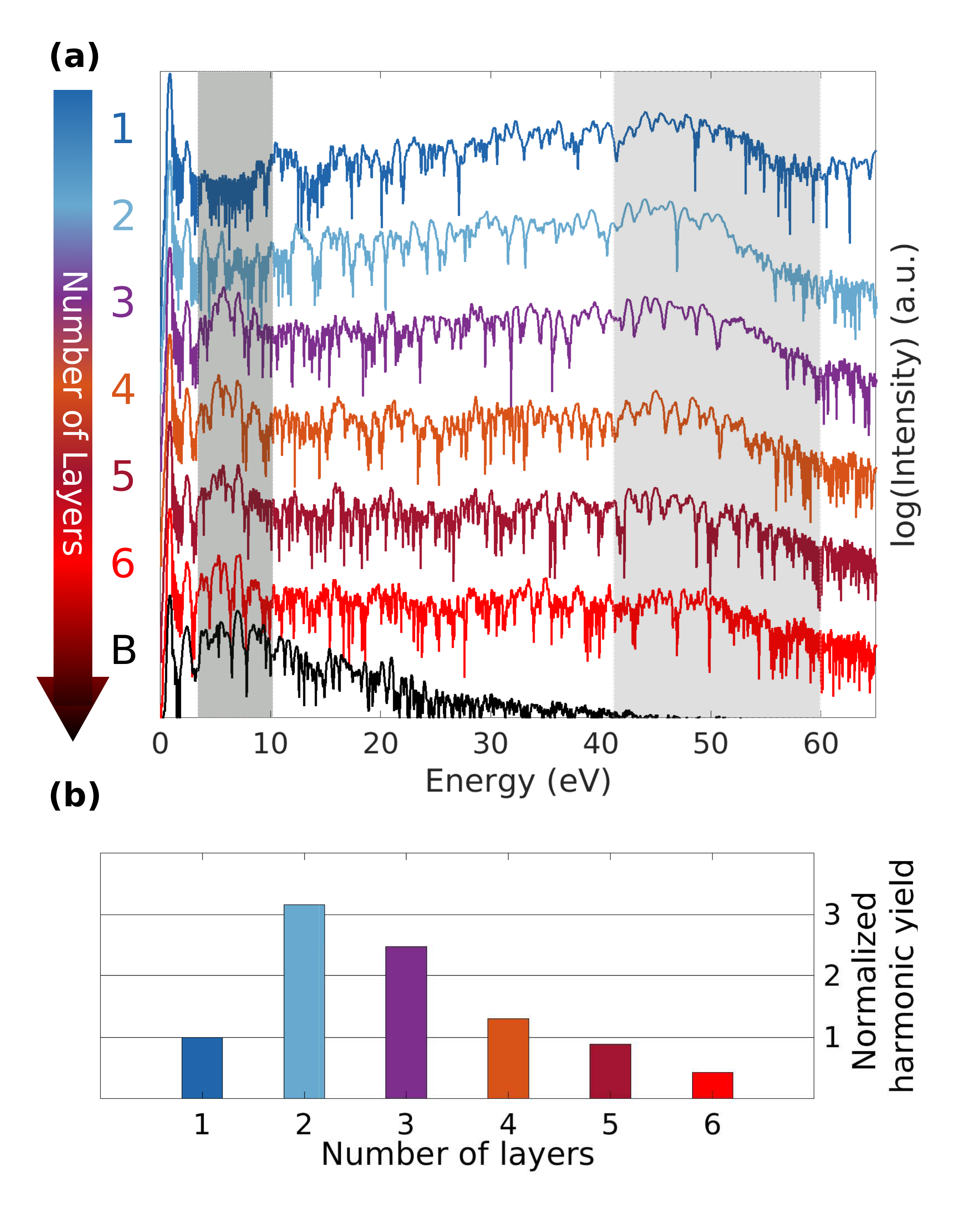}
    \caption{\label{fig:all_spectra} (a) HHG normalized spectra for one to six-layer slabs and the bulk one. The intensity in matter used for the bulk calculation is $6.85 \times 10^{12}$ W.cm$^{-2}$, as explained in the text. (b) Harmonic yield integrated  between 43 and 60 eV, normalized to the one of the monolayer, versus the number of layers. The shaded areas in (a) indicate the low-energy and high-energy regions, see the text for details.}
\end{figure}

The HHG spectra from hBN slabs composed of one to six layers are reported in Fig.~\ref{fig:all_spectra}(a), together with the bulk (black curve). The maximal intensity used for the slab systems is $5 \times 10^{13}$ W.cm$^{-2}$. The intensity for the bulk calculation ($6.85 \times 10^{12}$ W.cm$^{-2}$) has been chosen using the calculated induced electric field inside the slab as described below. We observe two main effects, one in the low-energy region, indicated by the dark-gray shaded area, and another one at higher energy (light-gray shaded region in Fig.~\ref{fig:all_spectra}(a)). Both effects are discussed in details below.

\subsubsection{Low-energy harmonics}
\begin{figure}
  \begin{center}
    \includegraphics[width=\columnwidth]{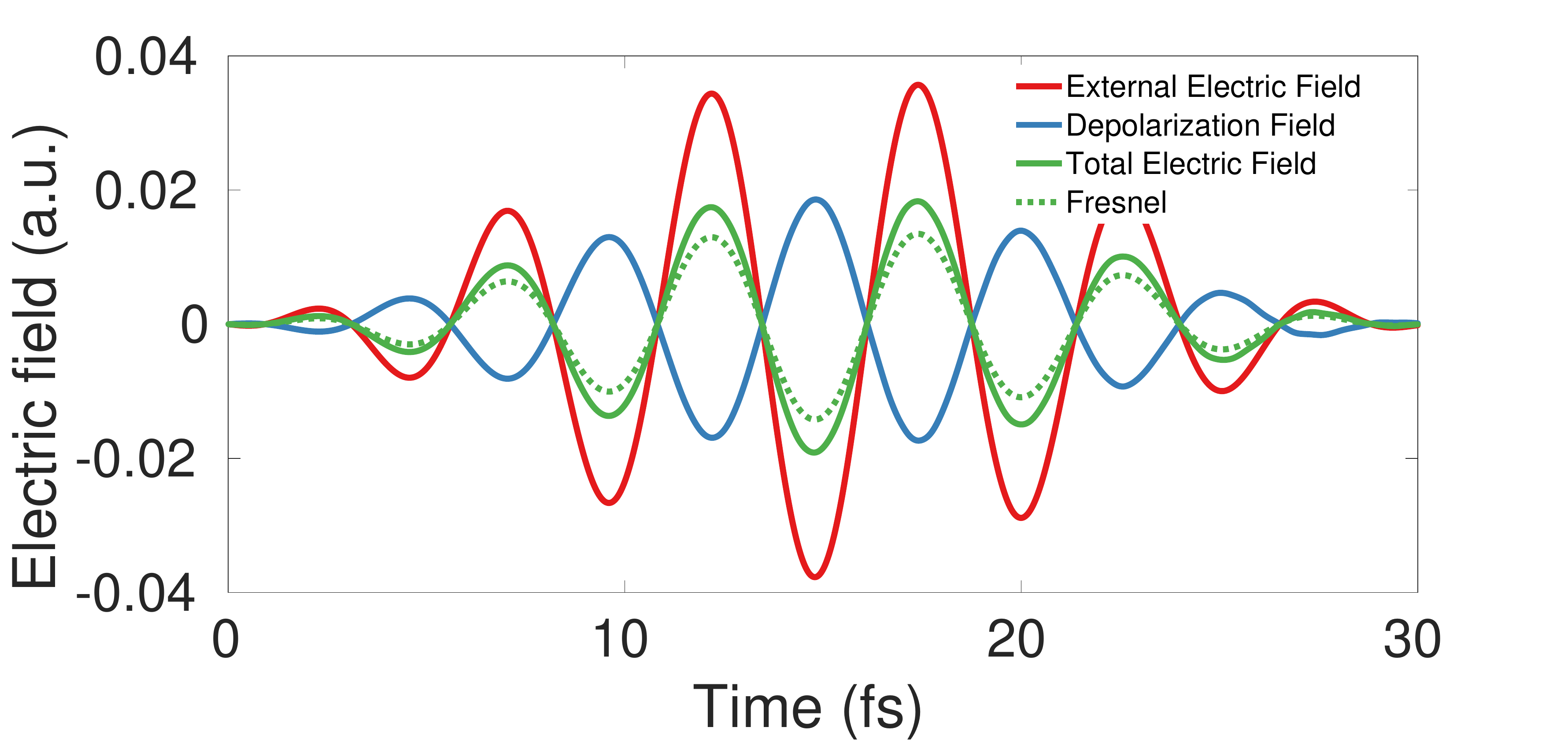}
  \end{center}
    \caption{\label{fig:external_induced_total} External electric field (red curve), surface induced electric field (blue curve) and total electric field (green curve) for the six-layer hBN slab driven by an out-of-plane electric field, see the main text for details. The green dashed curve shows the estimated total field from the Fresnel transmission coefficients, using the value of the bulk refractive index for a field polarized along the optical axis (n=2.25 ~\cite{h-BN_refractive_index}).
    }
\end{figure}

At low energy, we observe a clear increase of the spectral weight while increasing the number of layers. This increase of the spectral weight takes place in the spectral region corresponding to the harmonics from bulk hBN driven by a field polarized along the optical axis (corresponding to the out-of-plan direction).
These harmonics that grow while increasing the number of layers, therefore seem to correspond to the emergence of a bulk nature from the slabs. To confirm this, a precise comparison with the bulk is needed, which is the purpose of this section.\\
In order to precisely compare with the bulk HHG, and in particular the energy cutoff, one has to evaluate the intensity of the electric field acting on the electrons in the inner part of the slab, and to use it for computing the HHG from the bulk material.
One simple way to estimate it would be to use Fresnel coefficients, using either the experimental or the calculated optical refractive index of bulk hBN.
Here we decided to use instead a first principle approach, without assuming an abrupt interface, as done with Fresnel coefficients. The laser polarized along the direction perpendicular to the slab's surface creates an induced density oscillating at the frequency of the driving field (see Ref.~\onlinecite{tancogne-dejean_atomic-like_2018} for the monolayer case). This induced density is responsible for an induced electric field, as explained in Sec.~\ref{sec:method}B.\\
We computed this induced electric field, using Eq.\ref{eq:eind}, and extracted the induced field responsible for the screening of the external electric field due to surface charges. We refer to this induced field below as the ``depolarization field'', i.e., the field which is created in between the boundaries of the slab.
\begin{figure}
    \begin{center}
    \includegraphics[width=\columnwidth]{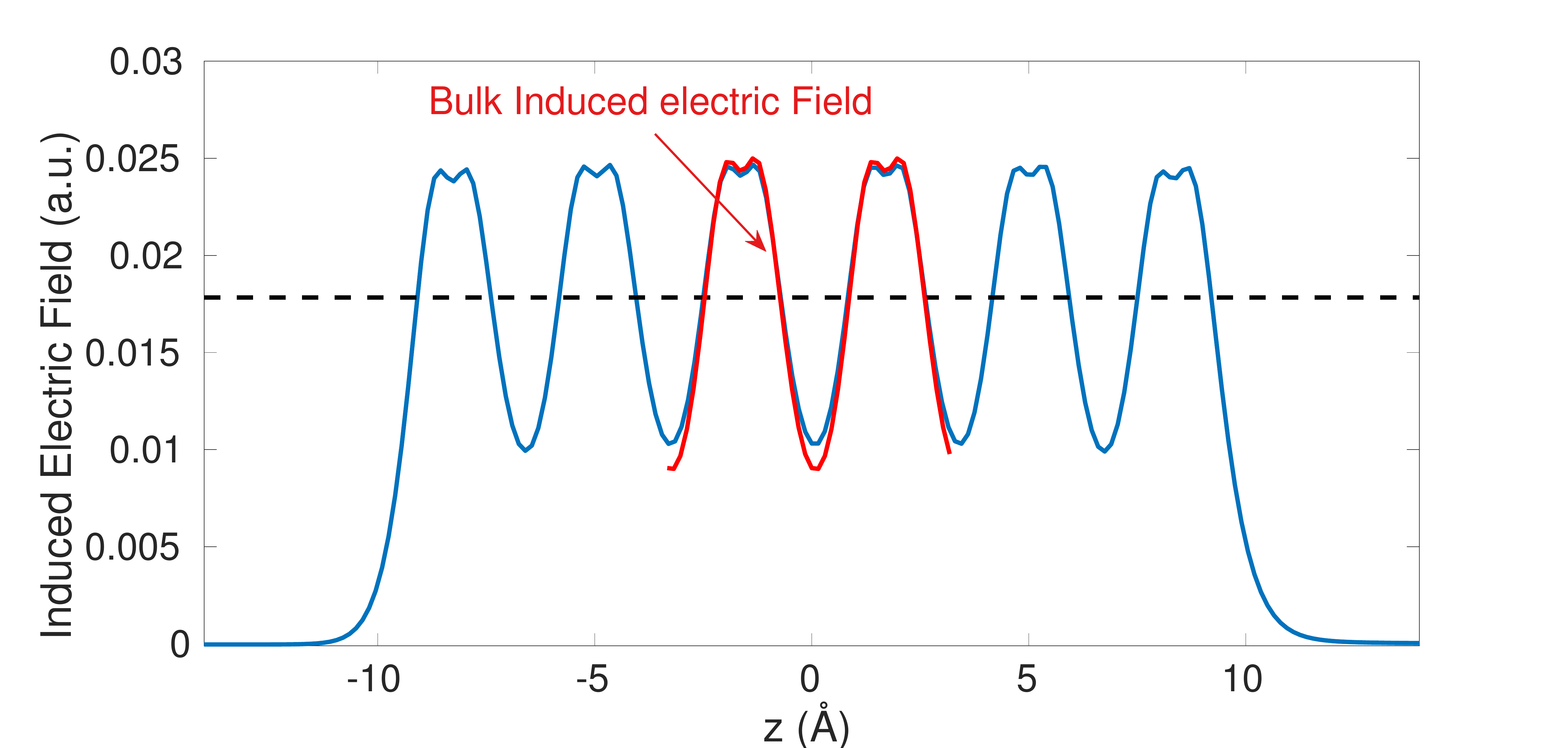}
  \end{center}
    \caption{\label{fig:induced_field} Induced electric field along the out-of-plane direction, averaged in the plane (blue curve). This induced electric field can be split into a surface contribution, which gives a uniform induced electric field, and a bulk contribution which, in average, gives no induced electric field. The bulk induced field (red curve) is shifted to the value of the depolarization field (dashed line), as explained in the text.}
\end{figure}
This is illustrated in Fig.~\ref{fig:induced_field}, in which we show the induced field along the out-of-plane direction computed at the maximal intensity for the six-layer hBN slab. In the case of a bulk material, the induced electric field (often referred as local fields) does not radiate, and therefore integrates to zero over space. In the region corresponding to the matter, we can therefore split the induced electric field shown in Fig.~\ref{fig:induced_field} into a spatially uniform part (the depolarization field), depicted by the black dashed line, and an oscillating part which integrates over zero. 
Taking the averaged value of the induced field between $z=\pm Nd_0/2$ ($d_0=6.29$ Bohr is the interlayer distance and $N$ is the number of layers) we computed the total electric field acting on the electrons inside the slab at each instant in time, as shown in Fig.~\ref{fig:external_induced_total}. \\
From this result, we can extract the value of the total electric field acting on the electrons in the central part of the slab, and use it to perform bulk calculations, without having to rely on Fresnel transmission coefficients.
To show the consistency of our approach, we reported in Fig.~\ref{fig:induced_field} (red curve) the induced electric field from the bulk, that we shifted by the value of the extracted depolarization field. It is clear that the induced electric field, and hence the induced density, behaves the same in the bulk and at the middle of the six-layer slab. This is a clear indication that already with a six-layer slab a bulk character is achieved inside the matter.
We also compared our result to what would be the total field assuming the Fresnel transmission coefficient, evaluated using the bulk refractive index (green dashed curve in Fig.~\ref{fig:external_induced_total}). Even if the results are similar, the agreement is not perfect, showing that using the bulk refractive index and the transmission coefficient is not a very precise procedure for few-layer systems.
\\
From our approach, we precisely estimated the total electric field acting on the electrons, fully taking into account the electric field induced by surface charges, without having to resort on Fresnel coefficients or a refractive index, which might not be valid in the case of intense driving fields.

\subsubsection{High-energy harmonics}

The evolution of the harmonic yield at high-energy while increasing the number of hBN layers in Fig.~\ref{fig:all_spectra}(b) is quite surprising. Assuming that the HHG from a few-layer system can be decomposed as ``surface'' and ``bulk'' contributions, one would expect that the ``surface'' generates atomic-like HHG whereas the ``bulk'' should gradually converge toward the true bulk HHG spectrum. 
We would therefore expect that the (unnormalized) harmonic yield for the highest harmonics (originating from the electrons been excited to the continuum, accelerated by the field and then recombined), should remain constant versus the number of layers. A change can be expected from a monolayer to the bilayer case, as these layers are stacked with different atoms facing each others (AA$'$ stacking, see Fig.~\ref{fig:view_system}(b)).

\begin{figure}
  \begin{center}
    \includegraphics[width=\columnwidth]{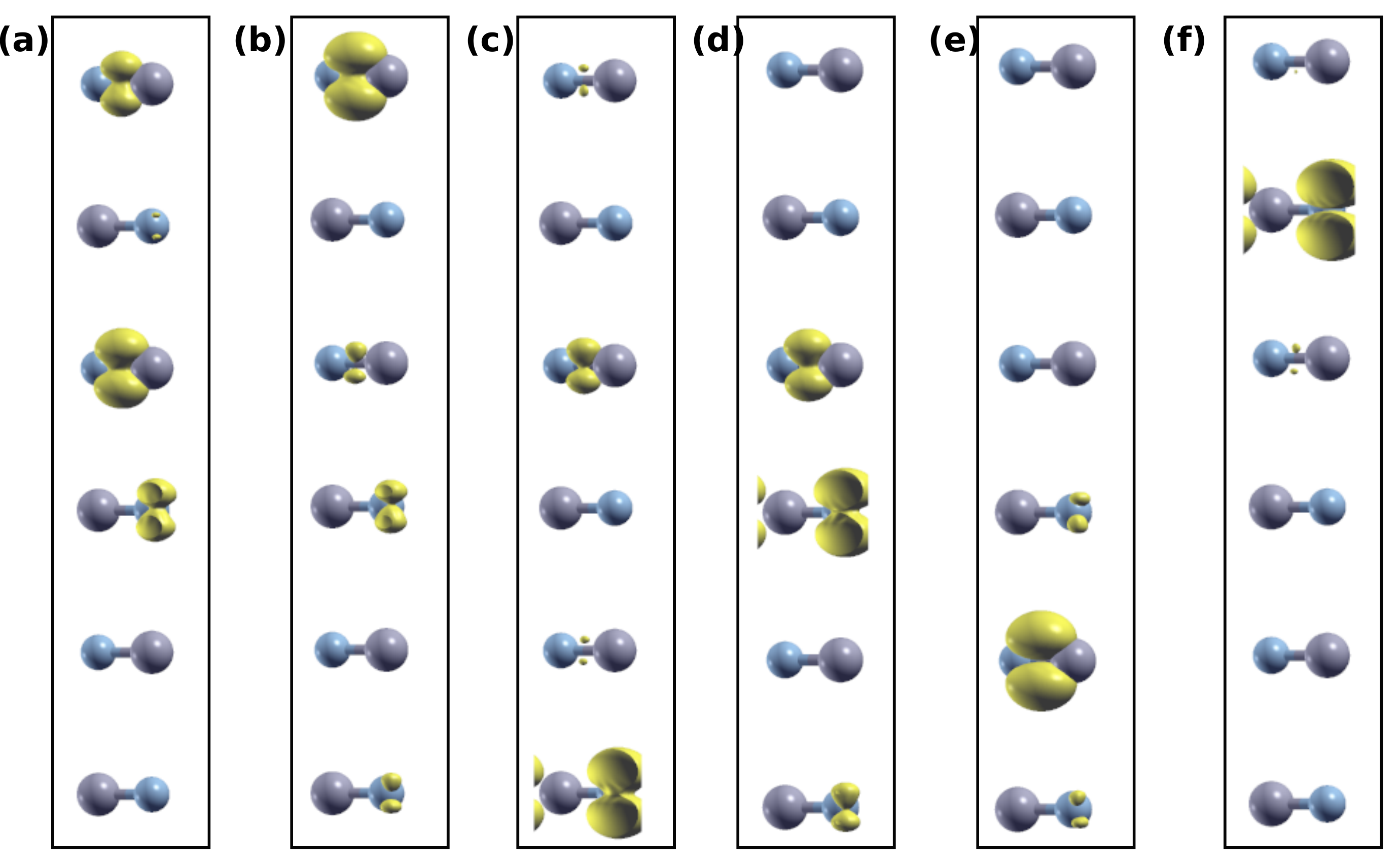}
  \end{center}
    \caption{\label{fig:wfn_6_layers} (a)-(f) Square modulus of the six highest occupied wavefunctions at the $\mathbf{k}$-point K, for the six-layer hBN slab. The value of the isosurfaces is taken here to be 0.08.}
\end{figure}

However, as shown in Fig.~\ref{fig:all_spectra}(b), this is quite not the case. Indeed, we observe that the harmonic yield, integrated between 43 eV and 60 eV,  first increases for the bilayer case, and then reduces  quickly as we increase the number of layers.
The increase from the monolayer to the bilayer can be expected as we double the number of electrons in the system, and hence we could expect up to a factor-of-four increase if there is no interaction between the layers. We found that this is almost the case.\\ 
When increasing more the number of layers, we found that the integrated yield quickly decreases.
From the previous analysis, we know that the inner layers feel a screened electric field, due to the depolarization effect.
We therefore do not expect these layers to contribute significantly to the atomic-like harmonic emission. The integrated yield should therefore remain more or less constant with an increasing number of layers.
The clear decrease of the integrated yield suggests that this simple picture is not completely true. The origin of the decrease is in fact understood with the help of Fig.~\ref{fig:wfn_6_layers}, which shows, in real space, the six highest occupied states at the K point of the Brillouin zone. The K point of the ground state corresponds to the location of the highest occupied electronic states.
The electrons are therefore excited to the continuum from the vicinity of this point (in reciprocal space), and the spatial extension of these wavefunctions will play a crucial role on both the ionization and the recombination of the electronic wavepacket responsible for the atomic-like harmonic emission \cite{tancogne-dejean_atomic-like_2018}. \\

\begin{figure}
  \begin{center}
    \includegraphics[width=\columnwidth]{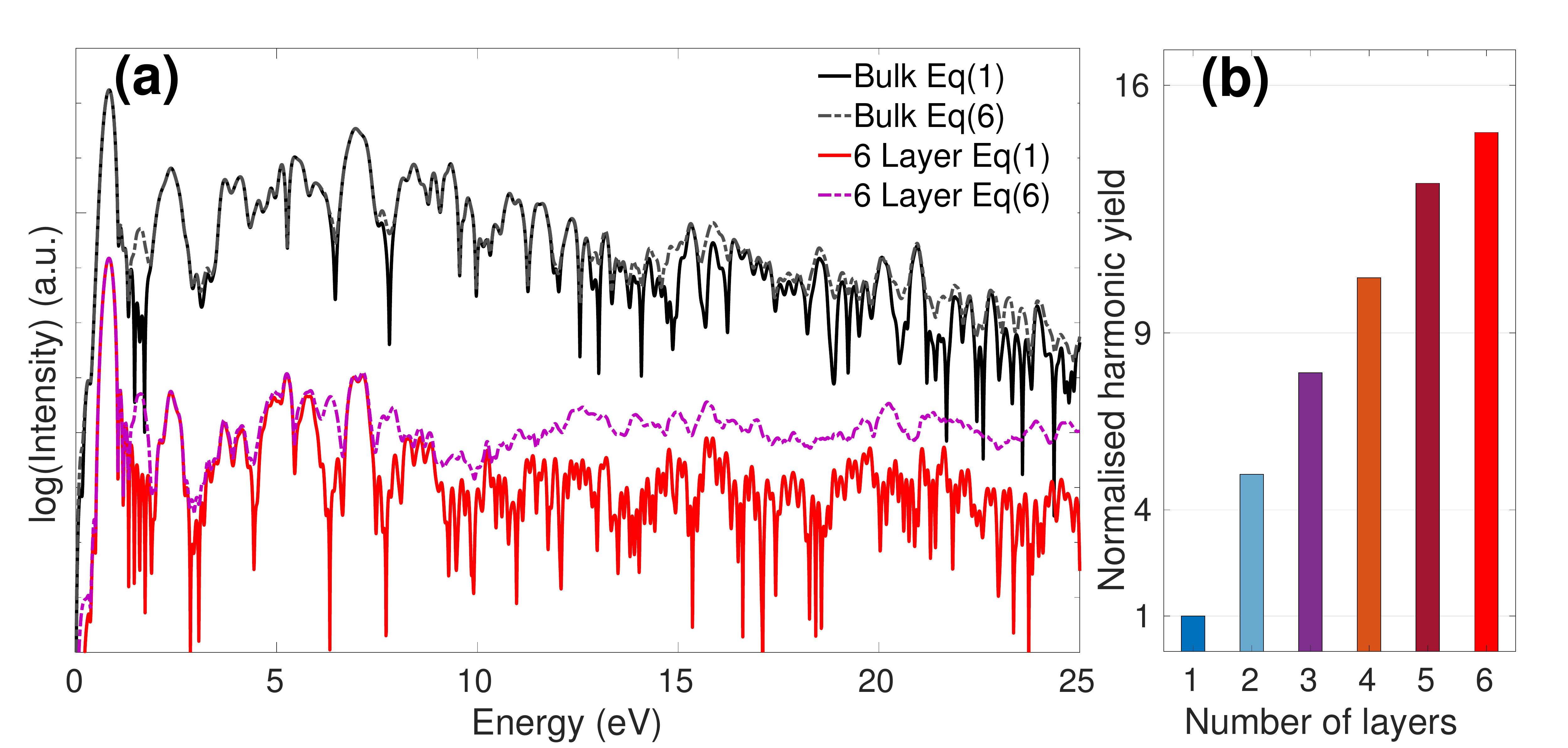}
  \end{center}
     \caption{\label{fig: effect of the phase along z high/low harmonics} Effect of the out-of-plan phase on the HHG spectra of the bulk hBN and a six-layer slab of hBN. (a) HHG spectra for the bulk and six-layer slab (black and red curves), and HHG spectra computed using Eq.~\ref{Eq: Spectra no phase formule}, i.e., without the out-of-plan phase (grey and purple dashed curves). (b) Same as Fig~\ref{fig:all_spectra}c, but using Eq.~\ref{Eq: Spectra no phase formule} to compute the harmonic spectra.}
\end{figure}

As one can see, none of the six highest occupied groundstate electronic wavefunctions at K are located at the edges of the six-layer slab of hBN. While these wavefunctions retain their $p_z$ nature, as in the monolayer, they are all clearly delocalized among the layers. This explains why the harmonic yield decreases. As we increase the number of layers, the wavefunctions start to delocalize in the out-of-plane direction, which corresponds to the emergence of a dispersion in the reciprocal space and the transition between a discrete level picture and dispersive electronic bands. Two reasons could be argued to explain a decrease of the yield. One would be that ionization is reduced by the delocalization of the wavefunctions, another one would be that the delocalization results in interferences between the different recombination channels that open, where one electron can leave from one layer and recombine to another one. We checked (not shown) by either integrating the electronic density outside of the matter part, or by looking at the number of electrons absorbed by the absorbing boundaries at the edges of the simulation box, that increasing the number of layers leads to more electrons been excited to the continuum. Therefore, we understand the reported decrease of the harmonic yield while increasing the number of layers as the result of destructive interferences between increasing number of possible quantum paths.\\
As a final check, we computed the quantity
\begin{equation}
\label{Eq: Spectra no phase formule}
I_{no phase}(\omega)= \Big( \int dz \left|\mathcal{F} \left( \int dx dy  \frac{\partial}{\partial t}\mathbf{j}(x,y,z,t)\right) \right| \Big)^2,
\end{equation} 
which corresponds to computing the HHG spectrum without taking into account the effect of the phase along the out-of-plane direction. Spectra computed using this formula are shown in Fig~\ref{fig: effect of the phase along z high/low harmonics}(a) (gray and purple curves) and the integrated high harmonics yield for each slab in Fig.~\ref{fig: effect of the phase along z high/low harmonics}(b).
If we use Eq.~\ref{Eq: Spectra no phase formule} instead of Eq.~\ref{Eq:hhg_formula}, more and more high order harmonics are generated as the number of layers is increased (see Fig.~\ref{fig: effect of the phase along z high/low harmonics}(b)), thus confirming our interpretation in terms of interference effects.
The simple picture of an atomic-like three-step-model mechanism, where electrons are ionized from one layer and recombine to this layer, quickly breaks when the extension of the system along the polarization direction of the laser field increases. Even a spatial extensions as small as 2 nm already shows strong modifications of the atomic-like picture, which cannot be used anymore to described such a system. Indeed, electrons are ionized from delocalized wavefunctions and recombine in the whole system. On the contrary, the low order harmonics are found to not be affected by the phase. In the bulk, the effect of the phase along the out-of-plane direction is almost negligible. For the six-layer slab, the low-order harmonics are also found to be very little affected by the phase along the out-of-plane direction before the calculated band-gap energy (4.5 eV) and more after. These interference effects therefore lead to a progressive destruction of the high-order harmonics originating from atomic-like trajectories, in benefit of the coherent, bulk-like, lower-order harmonics.

\begin{figure*}[t]
\begin{center}
    \includegraphics[width=0.42\textwidth, angle=-90]{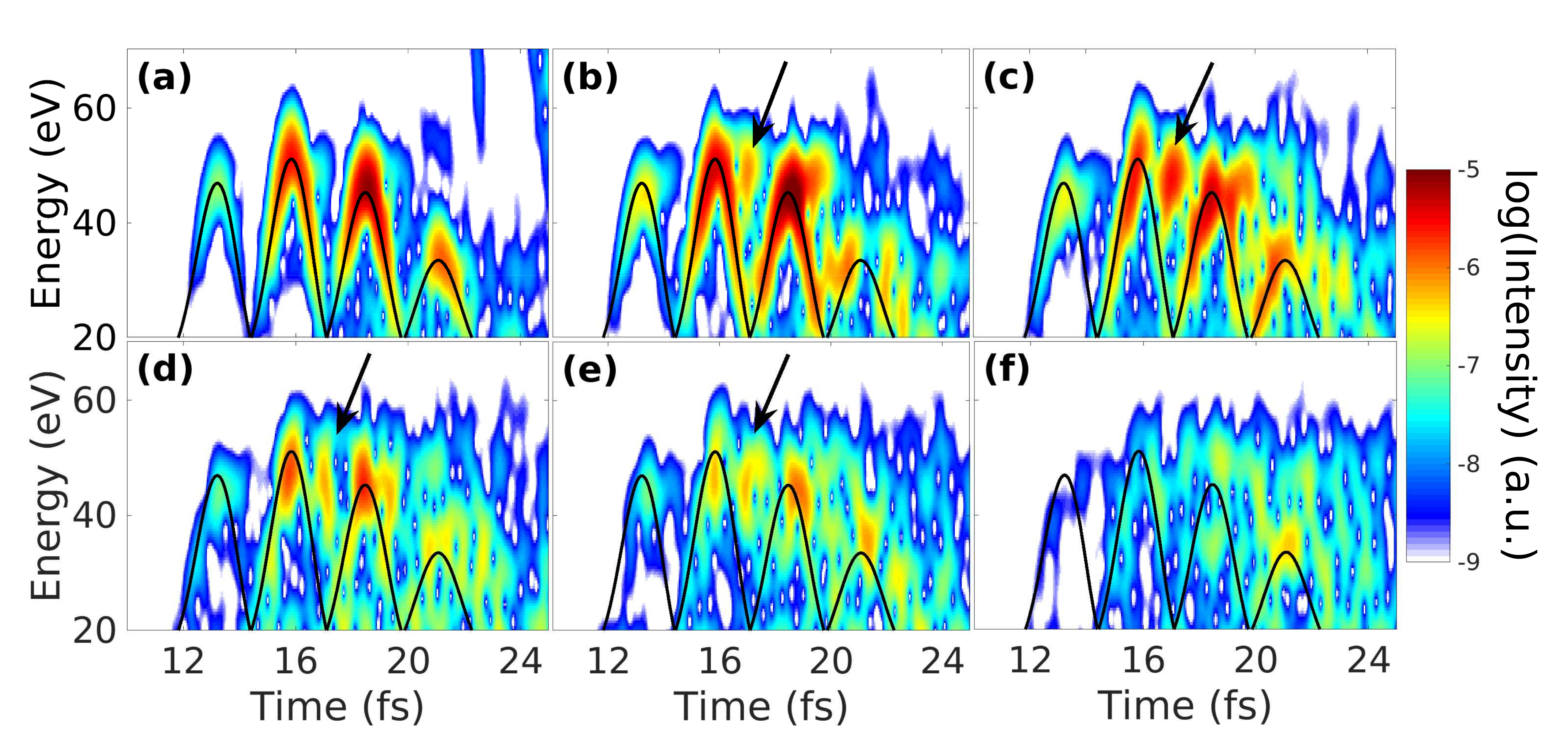}
  \end{center}
    \caption{\label{fig:gabor} Time-frequency analysis of the HHG from a monolayer of hBN (a) to the six-layer slab (f). A time window of 0.25 fs was used for computing the Gabor transforms. The black arrows in panels b-e show a secondary structure, which cannot be directly described by the three-step model. These structures appear as soon as two layers or more are stacked. The spectra become noisier as the number of layers stacked increases, because of the spatial interferences, as explain in the main text.}
\end{figure*}

The effect of the delocalization of the wavefunctions among the layers is also well visualized on the time-frequency analysis of the HHG spectra from one to six hBN layers. Our results reported in Fig.~\ref{fig:gabor} show that while increasing the number of layers, the clear trajectories progressively disappear, as a result of destructive interferences between electron wavepackets leaving and recombining at different layers in the systems. The semi-classical trajectories for the first return\cite{lewenstein_theory_1994} are shown in black, for which the values of each half-cycle peak field strength is used\cite{haworth_half-cycle_2007}.

We observe that the long trajectories are the first one to disappear. Moreover some side structures appear, for instance for bilayer hBN, as indicated by the arrow in Fig.~\ref{fig:gabor}(b). We checked that these structures disappear if we increase the distance between the layers (not shown), indicating that these structures result from the interaction of the two layers and not from the fact that two emitters (two layers) are included in the simulation. 

Our results have implications for the experimental observations of the predicted atomic-like HHG from two-dimensional materials~\cite{tancogne-dejean_atomic-like_2018}. Indeed, already for a six layers hBN slab, which corresponds to a sample thickness of 2 nm, the harmonic yield is reduced by almost a factor of two compared to the monolayer case. This indicates that a few-layer system very quickly recovers its bulk nature.
We note that this is in good agreement with a recent TDDFT study\cite{hansen_finite-system_2018} of 1D atomic chains, which found that the bulk limit is reached for chains longer than six atoms.\\
Moreover, our results show how the bulk response emerges as the result of the interference between the increasing number of quantum paths that open, as the wavefunctions delocalize though the entire system.

\section{Discussion}
\label{sec:discussion}

\begin{figure}[h!]
\begin{center}
    \includegraphics[width=0.46\textwidth, angle =-90]{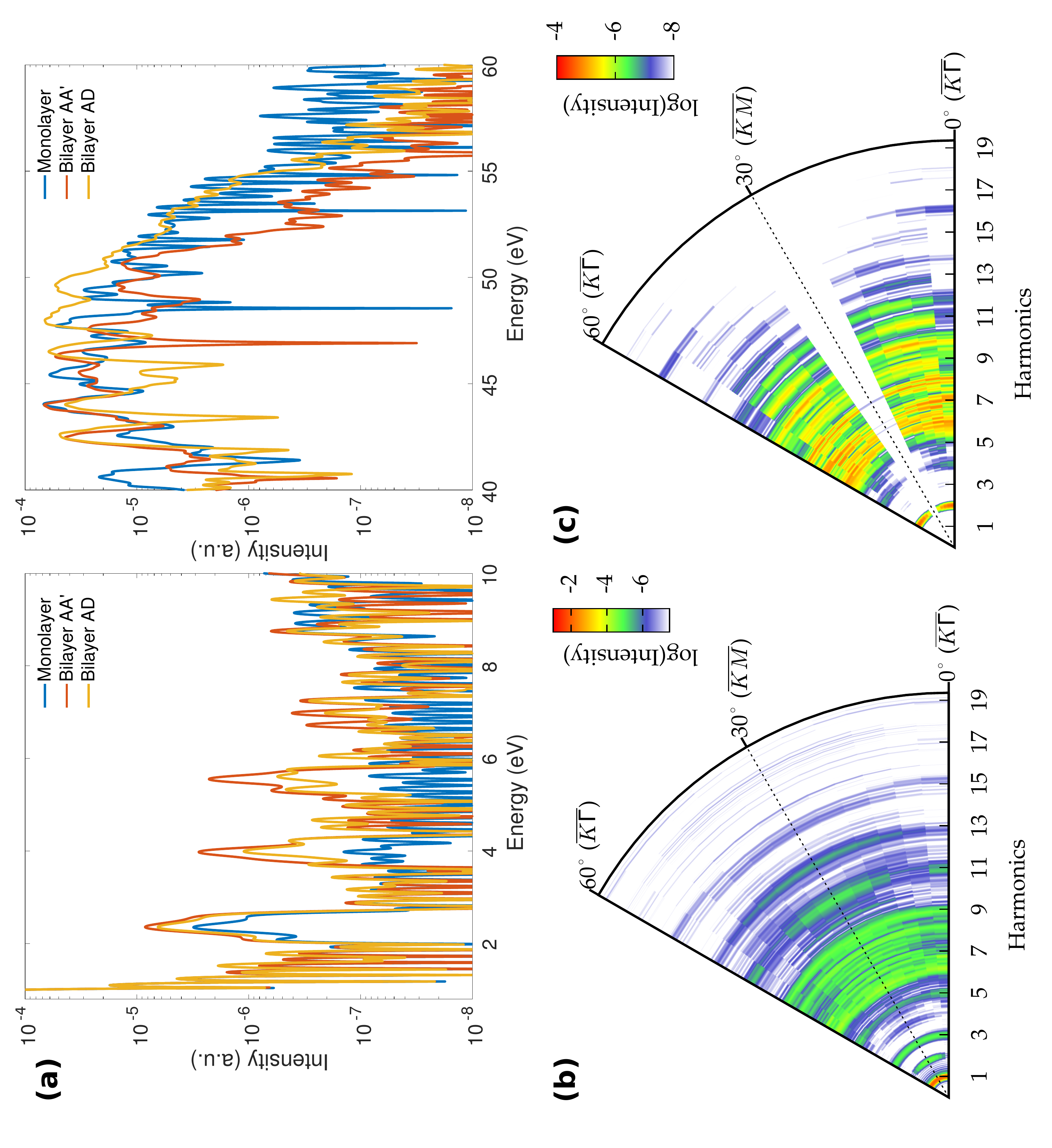}
  \end{center}
    \caption{\label{fig: inplane AA AD comparaison} (a) Atomic-like HHG spectra of the monolayer (blue curve), the bilayer with AA$'$ stacking (red curve), and the bilayer with AD stacking (orange curve). The left panel corresponds to the low-order harmonics whereas the right panel shows harmonics close to the energy cutoff. (b-c) Anisotropy of the harmonic emission for an in-plane polarized laser, respectively for the parallel (b) and perpendicular (c) contributions, for the bilayer with AD stacking.}
\end{figure}

While our main focus is the analysis of the evolution of the in-plane and out-of-plane HHG from few-layer systems with respect to the number of stacked layers, we also investigated some relevant aspects of the stacking configuration.\\
In Sec.~\ref{sec:results}, we showed that the number of stacked layers can influence the HHG emission. It is therefore natural to wonder how a different stacking could influence the harmonic emission. We decided to investing the so-called AD stacking (see Fig.~\ref{fig:view_system}), as this configuration has been shown to be also a stable stacking configuration~\cite{PhysRevLett.105.046801}.
For an in-plane laser polarization, the main difference with respect to the AA$'$ stacking is the generation of even harmonics, see Fig.~\ref{fig: inplane AA AD comparaison}(b-c). Indeed, in this two-layer system, inversion symmetry is broken, allowing for even harmonics to be emitted. 
The presence of even harmonics could therefore indicate not only a monolayer or a trilayer, but also a bilayer with a stacking different from the AA$'$ stacking. We found however that the intensity of the second harmonic scales with the number of layers in the AD stacking, whereas in the case of the AA$'$ stacking it is independent of the number of layers, as explained above.

In the case of a laser polarized along the out-of-plane direction, we found that the bilayer with AD stacking has the same induced electric field than the AA$'$ bilayer. This means that the depolarization field is very similar in both cases, and that the total electric field acting on electrons in not strongly affected by the stacking. However, we found that the atomic-like HHG from the AD stacking is quite different from the AA$'$ stacking. As shown in Fig.~\ref{fig: inplane AA AD comparaison}(a), the AD stacking leads to a lower harmonic yield than the AA$'$ stacking and a higher energy cutoff. The energy cutoff for the bilayer with AD stacking is in fact almost identical to the energy cutoff of the HHG spectrum of monolayer hBN. 
Moreover, we checked that the secondary structures that arise in the AA$'$ bilayer case in the time-frequency analysis (Fig.~\ref{fig:gabor}(b)) are less pronounced in the case of the AD stacking, confirming that these structures arise from the details of the interaction between the two layers.\\
Overall, these results indicate that the stacking of few-layer systems has an impact on the in-plane and out-of-plane HHG and modifies the electron dynamics in the strong-field regime.
We found that the AD stacking behaves very much like a monolayer, from both the point of view of the in-plane anisotropy, or looking at the energy cutoff of the atomic-like HHG spectrum. 
This might open up new directions of research, in which a specific stacking should be favored in order to improve the harmonic emission.

\section{Conclusions}
\label{sec:conclusions}

In conclusion, we have investigated the HHG from monolayer, few layers, and bulk hBN.
We focused on the effect of the layer stacking, and addressed the important question of how many layers does it take to obtain the bulk behavior.\\
Using extensive first-principle calculations within a time-dependent density functional theory framework, we showed how the in-plane and out-of-plane nonlinear non-perturbative response of two-dimensional materials evolves from the monolayer response to the bulk one.
Whereas the in-plane HHG is found not to be strongly altered by the stacking of the layers, we found that the out-of-plane response is strongly affected by the number of layers considered. This is explained by a combination of the induced electric field, resulting from the electron-electron interaction, and by a delocalization of the wavefunctions among the layers. 
We found that even if an increasing number of electrons are ionized, the resulting high-order part of the HHG spectrum progressively vanish and only lower-order harmonics, corresponding to the ones of the bulk, remain when the number of layer is increased.
We elucidated the transition of atomic-like harmonic emission to the bulk one as originating from destructive interferences between the contribution of the different layers, whereas the bulk contribution emerges as the remaining coherent part of the harmonic emission. \\
We briefly discussed the effect of the stacking for the case of bilayer hBN, showing that the type of stacking of a bilayer (AA$'$ stacking or AD stacking) does affect the harmonic emission, and in particular modifies the anisotropy of the in-plane harmonic emission for the bilayer case. This might open the door to the spectroscopy of few-layer systems using specific fingerprints of the stacking in the HHG spectrum.
\\
We believe that the present work will serve as a guideline for future experimental studies. Further studies should address the effect of the stacking in other vdW heterostructures, in particular with transition metal dichalcogenides, together with the role of excitonic effects, and of the electron-phonon coupling, two effects which are known to be crucial for electronic and linear optical properties of these materials.\\

\begin{acknowledgments}
We acknowledge financial support from the European Research Council (ERC-2015-AdG-694097), and Grupos Consolidados (IT578-13).
We would like to thank Oliver M\"ucke and Massimo Altarelli for fruitful discussions.
\end{acknowledgments}

\bibliographystyle{apsrev4-1}
\bibliography{biblio,2D_fewlayers}

\end{document}